\documentclass[genTeX]{nrc1}   

\volyear{??}{????}
\received{???}
\accepted{???}
\journal{Can. J. Phys.}

\newcommand{\beq}{\begin{equation}}
\newcommand{\eeq}{\end{equation}}
\newcommand{\bea}{\begin{eqnarray}}
\newcommand{\eea}{\end{eqnarray}}
\newcommand{\eq}[1]{(\ref{#1})}

\begin{document}

\title{g~factor in a light two body atomic system:\\
a determination of fundamental constants to test QED
}

\author{S. G. Karshenboim}
\address{D. I. Mendeleev Institute for Metrology, 198005 St. Petersburg, Russia and
Max-Planck-Institut f\"ur Quantenoptik, Garching, Germany}
\author{V. G. Ivanov}
\address{Pulkovo Observatory, 196140 St. Petersburg, Russia and
D. I. Mendeleev Institute for Metrology, 198005 St. Petersburg, Russia}

\shortauthor{Karshenboim and Ivanov}

\maketitle
\begin{abstract}
Energy levels of a two-body atomic system in an external homogeneous
magnetic field can be presented in terms of magnetic moments of their 
components, however, those magnetic moments being related to bound particles
differ from their free values.
Study of bound $g$~factors in simple atomic systems are now
of interest because of a recent progress in experiments
on medium $Z$ ions and of a new generation of muonium experiments
possible with upcoming intensive muon sources.
We consider bound corrections to the $g$ factors in several
atomic systems, experimental data for which are available in literature:
hydrogen, helium-3 ion, muonium, hydrogen-like ions
with spinless nuclei with medium $Z$.
\end{abstract}
\begin{resume}
French version of abstract (supplied by CJP)
   \traduit
\end{resume}

\section{Introduction}

High-resolution spectroscopy of hydrogen and hydrogen-like atoms provides significant amount of data for precision QED tests and
determination of basic fundamental physical constants such as the {\em
Rydberg constant} $R_\infty$, the {\em fine structure constant} $\alpha$,
etc. However, not only the spectrum of a free hydrogen-like atom can be
studied, but also its interaction with an external field and in particular
with a homogeneous magnetic field. From experimental and practical point
of view two kinds of two-body systems are of particular interest. Those
are {\em muonium} and an {\em hydrogen-like ion with a spinless nucleus} at not
too high $Z$.

$g$~factor of a bound lepton (electron or muon) can be presented in the form
\cite{sgk00}
\beq\label{1ab}
g = 2 \cdot \big(1+a+b\big)\:,
\eeq
where {\em 1} stands for the Dirac value, $a$ is anomalous magnetic moment
of a free lepton and an additional correction $b$ is due to the binding
effects. The problems for calculation of $b$ in muonium and an hydrogen-like ion are
similar but they have some different significant features.

\section{Motivation: muonium}

Properties of the hydrogen atom are affected by effects of the proton
structure which limit accuracy of any theoretical prediction. The muonium
atom is, in contrast to hydrogen, a pure leptonic system suited for an
accurate QED test. A value of its hyperfine structure has been measured at
LAMPF with a great accuracy
\cite{MuExp}
\beq \label{MuHhsExp}
\nu_{\rm hfs} ({\rm exp}) = 4\,463\,302.776(51)~\mbox{kHz}\,.
\eeq
The experimental uncertainty of $\nu_{\rm hfs}$ is only $1.1\cdot10^{-8}$ in fractional units.
The calculation has not been able to reach this level yet. The theoretical problems are:
\begin{itemize}
\item
It is not possible to do any exact calculations. One have to expand either
in $\alpha$ (i.e. to select diagrams with a different number of QED loops), or 
$Z\alpha$ (i.e. over the binding effects) or $m/M$ (i.e. over recoil
effects). A possible contribution of presentely uncalculated terms of higher order
has been estimated as $5\cdot 10^{-8}$ \cite{cek,ZP96}.
\item
One has to take into account hadronic effects (see e.g. \cite{cek,ekscjp}).
\item
It appears that a largest problem of the {\em theoretical} calculation is
an {\em experimental} uncertainty. The leading contribution into the ground state
hyperfine splitting (so-called Fermi energy)
\beq\label{EFermi}
\nu_F=\frac{16}{3}{(Z\alpha)}^2 Z^2R_{\infty} c \frac{m_e}{m_\mu}
\left [ \frac{m_\mu}{m_\mu+m_e}\right ]^{3}(1 + a_{\mu} )
\eeq
contains the electron-to-muon mass ratio ${m_e}/{m_\mu}$. The most accurate determination
of the mass ratio has been performed at LAMPF \cite{MuExp} studying Breit-Rabi levels
of the muonium ground state in presence of homogenous magnetic field. For any successful
interpretation of the experiment, $g$~factors of muon and electron in muonium are needed.
\end{itemize}

Special features of the theory of $g$~factors of electron and muon in muonium are:
\begin{itemize}
\item
It is necessary to find a theoretical expression for the bound $g$~factors with an uncertainty below
0.01 ppm.
\item
All small parameters, $\alpha$, $Z\alpha$ and ${m_e}/{m_\mu}$, are of about the same value
($\alpha=Z\alpha\sim 1/137$ and ${m_e}/{m_\mu}$)
and hence QED, binding and recoil effects are
equally important.
\end{itemize}
The theory up to the {\em third order} in either of these three small parameters has been known up to now and our
target is a complete evaluation of {\em fourth-order} corrections.

\section{Motivation: Hydrogen-like ions with spinless nuclei}

Recent Mainz-GSI experiment \cite{werth,werth01,werthcjp}
have provided precision measurement of
\begin{itemize}
\item ion cyclotron frequency
\beq
\omega_{ion} = \frac{(Z-1) e}{M_{ion}}\,B
\eeq
\item and spin precession frequency
\beq
\omega_{spin} = g_e^\prime\frac{e}{2m_e} B
\eeq
\end{itemize}
with uncertainty below 1~ppb. A comparison of these two
frequencies leaves us with two values: of the $g$~factor of a bound
electron and of the electron-to-ion mass ratio. Accuracy of the latter is
essentially the same as of determination of the electron-to-proton mass ratio in
Ref.~\cite{farnham95} and in fact 
the uncertainty for $m_e/M_{ion}$ derived from Ref.~\cite{farnham95} is bigger than both the experimental
uncertainty for frequencies and the theoretical uncertainty for the bound
$g$~factor (see e.g. \cite{sgkH2}). The comparison of experiment with theory provides the most
accurate determination of the electron-to-proton mass ratio and a test of
QED calculations. The latter is possible if the experiment on two
different ions is performed and hence a ratio of their $g$~factors is free of any
uncertainty due to the proton-to-electron mass ratio. Published
experimental data are related to the ion of carbon-12 \cite{werth}, recent efforts has 
been directed to the oxygen-16 ion \cite{werthcjp}. Precision measurements with
other ions are also possible and expected.

The theory must target a level of accuracy as good as few parts in $10^{10}$.
The details of theory differ from muonium: the recoil effects are pretty small
and less important than the QED contributions. The most important problem is to
take into account the binding
effects. The latter are less important for lighter H-like ions, namely for
hydrogen-like ions
of {\em helium-4} and {\em beryllium-10}.

\section{Muonium: calculations}

The ``old'' theoretical expressions for the ground state in muonium ({\em up to the third
order}) are of the form (see e.g. \cite{th})
\bea
 g_e^\prime
({\rm up~to~3rd})
 &=&
 g_e^{(0)}\cdot\left\{1-{(Z\alpha)^2\over3}
 \left[1-{3\over2}{m_e\over m_\mu}\right]
 +{\alpha(Z\alpha)^2\over{4\pi}}
 \right\}
\eea
and
\bea
 g_\mu^\prime({\rm up~to~3rd})
 &=&
 g_\mu^{(0)}\cdot
 \left\{1-{\alpha(Z\alpha)\over3}
 \left[1-{3\over2}{m_e\over m_\mu}\right]
 \right\}
  \,.
\eea
Although $Z=1$ for muonium we keep it to simplify classification of contributions.

We present here a complete result for the {\em fourth order} corrections
\bea \label{fourthe}
 g_e^\prime({\rm 4th})
 &=&
 g_e^{(0)}\cdot\Biggl\{
 -{(Z\alpha)^2(1+Z)\over2}\left({m_e\over m_\mu}\right)^2
 -{5\alpha(Z\alpha)^2\over{12\pi}}{m_e\over m_\mu}
 \nonumber\\
 &&
 -{(Z\alpha)^4\over12}
 -\bigl(0.289\dots\bigr)\cdot\frac{\alpha^2(Z\alpha)^2}{\pi^2}
 \Biggr\}
\eea
and
\beq \label{fourthmu}
  g_\mu^\prime({\rm 4th})
=
  g_\mu^{(0)}\cdot
  \Biggl\{
 -{\alpha(Z\alpha)(1+Z)\over2}\left({m_e\over m_\mu}\right)^2+
 {\alpha^2(Z\alpha)\over{6\pi}}{m_e\over m_\mu}
 -{97\over108}\,\alpha(Z \alpha)^3
 \Biggr\}
\,.
\eeq
Most of terms in Eqs.~\eq{fourthe}, \eq{fourthmu} are of the {\em kinematic} origin and we have systematically checked
them. All of those kinematic terms can be derived by several methods applied originally for derivation of the {\em third
order corrections} which are all of the {\em kinematic} origin (cf. \cite{th}). Non-{\em kinematic} corrections include a
high-order Breit contribution to $g_e^\prime$ \cite{breit28} and a screening contribution to $g_\mu^\prime$. The latter was calculated using of the corrections to the Dirac wave function due to hyperfine interaction \cite{jetp00} and due to 
external magnetic field \cite{jetp01}. The result agrees with calculations for the screening tensor in Refs.~\cite{pyper97,moore97}.

The results of a re-evaluation of LAMPF data \cite{MuExp1,MuExp} taking into account Eqs.~\eq{fourthe}, \eq{fourthmu} is shown in Table~\ref{tab1}.

\begin{table}
\caption{Re-evaluation of LAMPF data. The fractional corrections $\delta_{e,\mu}$ to the derived
value of the ratio of the muon magnetic moment and proton magnetic moment
are related to the fourth order contribution to $g_{e,\mu}$ found in this work.}
\label{tab1}
\begin{center}
\begin{tabular}{ccc}
\hline*
Experiment & LAMPF, 1982, \cite{MuExp1} & LAMPF, 1999, \cite{MuExp}\\
\hline
$\mu_\mu^\prime/\mu_p$& {~$3.183\,290\,0(13)$} & {~$3.183\,289\,03(38)$} \\
$\mu_\mu/\mu_p$ & {~$3.183\,346\,1(13)$} & {~$3.183\,345\,14(38)$} \\
\hline
$\delta_e$ & {$-0.4\cdot10^{-8}$} & {$-0.3\cdot10^{-8}$} \\
$\delta_\mu$ & {$\phantom{-}1.2\cdot10^{-8}$} &
{$\phantom{-}1.2\cdot10^{-8}$} \\
\hline*
\end{tabular}
\end{center}
\end{table}

\section{H-like ions with spinless nuclei: calculations}

We present a theoretical expression for the binding correction to the electron $g$ factor in the form
\beq
b= b_{\rm kin} + b_{\rm rel} + b_{\rm rec} + b_{\rm rad} + b_{\rm nucl}\,.
\eeq
The {\em kinematic} term was discussed above in the case of muonium. The
{\em relativistic} corrections are known exactly thanks to
Breit \cite{breit28}.
The {\em nuclear corrections} were considered in
\cite{beier,sgk00,glazov01}. For $Z<10$
an analytic expression \cite{sgk00}
\beq
b_{\rm nucl} =\frac{4}{3}\,(Z \alpha )^4 \,(mR_N)^2
\eeq
has an adequate accuracy.

The {\em recoil} effects are small enough. We have
performed an evaluation based on the Grotch equation and confirm here the
leading term
\beq \label{dbscm}
  b_{\rm rec} =
  -\frac{1}{24}(Z\alpha)^4 \frac{m}{M}
  \,,
\eeq
obtained recently in Ref.~\cite{yelk01}. There were also numerical
calculations by Shabaev and Yerokhin \cite{shab01} and we apply 
their data to obtain our final results.

The essential binding effects contribute into the {\em radiative}
corrections
\beq
  b_{\rm rad} = b_{\rm VP}+ b_{\rm WK}+ b_{\rm SE}+b_{\rm 2loop} \,.
\eeq
The {\em vacuum polarization} effects were studied in Ref.~\cite{beier} numerically and in Ref.~\cite{jetp01} analytically. The {\em Wichman-Kroll} contribution was investigated in Ref.~\cite{sgk00}. The other light-by-light contributions are needed more study \cite{progress}. The {\em two-loop} contributions have not been yet calculated and only some conservative estimations are possible. The biggest contribution and the biggest uncertainty in $b_{\rm rad}$ comes from the one-loop electron {\em self energy}. Fractional uncertainty of numerical data for $b_{\rm SE}(Z)$ encreases at lower $Z$ and one can impove accuracy of low $Z$ results after properly fitting data. We fit here numerical data from Ref.~\cite{beier}. An essential improvement for $Z=2$ and $Z=4$ has been achieved. After our analysis has been completed a paper \cite{oneloop} with data more accurate than thoes in Ref.~\cite{beier}. Results of our fitting are in fair agreement with new numerical calculations.

The final results for ions of interest are collected in Table~\ref{tab2}
below. Note, that the theoretical accuracy for the hydrogen-like ions of
helium-4 and beryllium-10
is higher than for carbon and oxygen and from theoretical point
of view they are better suited for determination of the electron-to-proton
mass ratio.

\begin{table}
\caption{The binding contributions to $g$~factors of some hydrogen-like ions.}
\label{tab2}
\begin{center}
\begin{tabular}{cr}
\hline*
Ion & \multicolumn{1}{c}{$b$}\\
& \multicolumn{1}{c}{[$10^{-9}$]}\\
\hline
$^4$He$^+$ & $-70\,948.84(3)$\\
$^{10}$Be$^{3+}$ & $-283\,865.0(2)$\\
$^{12}$C$^{5+}$ &  $-638\,857.4(5)$\\
$^{16}$O$^{7+}$ & $-1\,136\,142.5(8)$ \\
$^{18}$O$^{7+}$ & $-1\,136\,141.9(8)$\\
\hline*
\end{tabular}
\end{center}
\end{table}

A comparison of carbon-to-oxygen and oxygen-to-oxygen leads to
\beq\label{gthratio}
g(^{12}{\rm C} ^{5+}) /g(^{16}{\rm O} ^{7+})
=1.000\,497\,273\,4(9)
\,,
\eeq
and
\beq\label{gOO}
  g(^{16}{\rm O} ^{7+})-g(^{18}{\rm O} ^{7+})
  =
  -1.2\cdot10^{-9}
  \,.
\eeq
Equation \eq{gthratio} can be used to test QED calculations, while Eq.~\eq{gOO} is rather useful to check for experimental systematical errors. The experimental result for the ratio in Eq.~\eq{gthratio} is
\beq
g(^{12}{\rm C} ^{5+}) /g(^{16}{\rm O} ^{7+})
=1.000\,497\,273\,1(15)\,,
\eeq
where we take into account a possible correlation between systematic errors for the carbon experiment 
\cite{werth} and the oxygen measurement \cite{werthcjp}. The experimental value is in a fair agreement with the theoretical prediction \eq{gthratio}.

\section{Muon-to-electron mass ratio}

Several results of determination of the muon-to-electron mass ratio are
collected in Table~\ref{tab3}.
The improvement of theory for the $g$~factors of a bound electron and muon in
the muonium atom is not that important for present level of accuracy. However, a number
of promising projects for intensive muon sources for experiments in particle
physics are now under consideration \cite{sources}. The accurate theory of the $g$~factors
in muonium will be useful for upcoming experiments.

\begin{table}
\caption{Muon-to-electron mass ratio. The results marked with
$\dag$ differ from originally published. The result of Beltrami et al. is
corrected accordingly to Ref.~\cite{ZP96}. The result of Casperson et al.
is presented accordingly to Ref.~\cite{klempt}. The results marked with
$\dag$ are found with taking into account fourth order corrections to the
bound $g$ factors of electron and muon. The CODATA result is based on some result derived from {\em 1s
HFS Mu} with some overoptimistic estimation of theoretical uncertainty. The result from {\em 1s
HFS Mu} quoted here is calculated with estimation of theoretical uncertainty as
explained in Ref.~\cite{cek} and with use of $1/\alpha
=137.035\,999\,58(52)$ from anomalous magnetic moment of electron
\cite{kinoshita,trap}. }
\label{tab3}
\begin{center}
\begin{tabular}{lll}
\hline*
\multicolumn{1}{c}{Method}
& \multicolumn{1}{c}{$m_\mu/m_e$}
& \multicolumn{1}{c}{Ref.}\\
\hline
Precession of $\mu^+$ in water &206.768\,60(29)
& Crow et al., 1972, \cite{Crow} \\
Precession of $\mu^+$ in Br$_2$
&206.768\,35(11)
& Klempt et al., 1982, \cite{klempt} \\
$(g-2)$ of $\mu^-$ & 206.771\,4(21)
& Bailey et al., 1977, \cite{Bail77}  \\
$(g-2)$ of $\mu^+$ & 206.766\,8(20)
&  Bailey et al., 1977, \cite{Bail79} \\
$(g-2)$ of $\mu^+$ & 206.768\,70(27)
&  Brown et al., 2001, \cite{amuexp,lbl} \\
$3d_{5/2}-2p_{3/2}$ transitions & & \\
in muonic $^{24}$Mg and $^{28}$Si
&206.767\,94(64)$^\star$ & Beltrami et al., 1986, \cite{Belt} \\
Breit-Rabi for Mu & 206.768\,18(54)$^\star$ & Casperson et al., 1977, \cite{Casp77}  \\
Breit-Rabi for Mu & 206.768 22(8)$^\dagger$
& Mariam et al., 1982, \cite{MuExp1} \\
Breit-Rabi for Mu & 206.768\,283(25)$^\dagger$
& Liu et al., 1999, \cite{MuExp} \\
1s HFS Mu & 206.768\,283(11)$^\star$ &Liu et al., 1999, \cite{MuExp} \\
$1s-2s$ in Mu  & 206.768 38(17)
& Meyer et al., 2000, \cite{meyer00} \\
\hline
Adjustment& 206.768\,265\,7(63)& CODATA, 1998, \cite{codata} \\
\hline*
\end{tabular}
\end{center}
\end{table}

\section{Proton-to-electron mass ratio}

Determination of the proton-to-electron mass ratio is summarized in
Table~\ref{tab4} below.
Presently the most accurate value of the electron-to-proton mass ratio
comes from an experiment related to the $g$~factor of a bound electron.
Theory and experiment contribute into uncertainty at the same level. Experiments with lighter ions might reduce the theoretical uncertainty substantially.

\begin{table}
\caption{Proton-to-electron mass ratio. The result for $^9$Be$^+$ differs
from the original value in Ref.~\cite{wineland} because we have applied a
more accurate theory from Ref.~\cite{yan} and a more accurate value of the
beryllium-9 mass \cite{firestone}. We present two theoretical evaluation
of experiment \cite{werth}. The value marked with $\dag$ has been obtained
in this work.}
\label{tab4}
\begin{center}
\begin{tabular}{lll}
\hline*
\multicolumn{1}{c}{Method}
& \multicolumn{1}{c}{$m_p/m_e$}
& \multicolumn{1}{c}{Ref.}\\
\hline
$g$~factor in $^{9}$Be$^{+}$ & 1\,836.152\,92(29)$^\star$ &Wineland et al., 1983, \cite{wineland}\\
$g$~factor in $^{12}$C$^{5+}$ & 1\,836.152\,673\,3(14) & H\"affner et al., 2000, \cite{werth}\\
& & Beier et al., 2002, \cite{PRL}\\
$g$~factor in $^{12}$C$^{5+}$ & 1\,836.152\,673\,6(13)$^\dagger$
& H\"affner et al., 2000, \cite{werth}\\
Precession of $e$ and $p$ & 1\,836.152\,667\,0(39) & Farnham et al., 1995, \cite{farnham95} \\
Precession of $e$ and  H$_2^-$ & 1\,836.152\,680(88) & Gabrielse et al., 1990, \cite{HminusE} \\
Gross structure in H and D & 1\,836.152\,667(85)& de Beauvoir et al., 2000, \cite{paris}, \\
& & Huber et al., 1998, \cite{mpq}  \\
Antiprotonic He & 1\,836.152\,58(24) & Hori et al., 2001, \cite{hori01} \\
\hline
Adjustment & 1\,836.152\,667\,5(39)& CODATA, 1998, \cite{codata} \\
\hline*
\end{tabular}
\end{center}
\end{table}

\section{The fine structure constant}

The mass ratios of electron-to-muon and electron-to-proton are considered
above. Both are important for precision determination of the fine
structure constant $\alpha$. A precision value of the latter is strongly
needed for any accurate QED
test.

Comparison of theory and experiment for the muonium hyperfine structure interval
allows to determine $\alpha$ if a value of the muon mass is known in a proper units, e.g in unit of the electron mass, or the muon magnetic moment in unit of the magnetic moment of electron or proton.

A promising method to determine the fine structure constant has been
developing by Chu and coworkers \cite{chu}. It is based on the photon recoil spectroscopy
and offers a precision value of $h/M_{\rm atom}$. The
experiment in Ref.~\cite{chu} deals with cesium atoms, but some other atoms are also used in
several experiments by other groups. order to obtain precise value for $\alpha$, one can combine $h/M_{\rm atom}$ with a precision value of the {\em Rydberg constant}
\beq
R_\infty ={\alpha^2\over 2} \, \frac{m_ec}{h}\;,
\eeq
the cesium-to-proton mass \cite{rain} and the electron-to-proton mass ratio.

The presently most accurate value of the fine structure constants comes
from study of anomalous magnetic moment of electron \cite{kinoshita}. It used to be quoted
as a {\em QED} value. However, the major uncertainty of this value
of $\alpha$ comes from not perfect understanding of a motion of a light particle (i.e.
electron) in a {\em Penning trap} \cite{trap}. It is important that the value of the
electron-to-proton mass ratio from the bound $g$~factor is free of problem
of an electron in Penning trap in contrast to the former best value of the
ratio from University of Washington \cite{trap}. That makes a value of $\alpha$ from
Chu experiment really independent from those from $g\!-\!2$ of
electron. The recent preliminary result by Chu has a relative uncertainty as low as $7\cdot 10^{-9}$
and some progress is still possible \cite{chu}.

\section*{Acknowledgements}

The authors would like to thank G\"unther Werth and all Mainz-GSI team,
Klaus Jungmann, Rudolf Faustov, Peter Mohr, Alexander Milstein and Vladimir Shabaev for stimulating
discussions. The work was supported in part by RFBR grant 00-02-16718.

\end{document}